\documentclass[12pt]{article}
\pdfoutput=1

\usepackage{amsmath,amssymb}
\usepackage{epsfig} 
\usepackage{graphicx, rotating}
\usepackage{hyperref}
\usepackage{slashed}
\usepackage{amsmath}
\usepackage{slashed}
\usepackage{cite}
\usepackage{multicol}
\usepackage{color}

\newcount\Mac  \Mac=1  
\newcommand{\ifMac}[2]{\ifnum\Mac=1 #1 \else #2 \fi}
\def\putps(#1,#2)(#3,#4)#5#6{\ifnum\Mac=1 \put(#1,#2){\special{picture #5}}
\else  \put(#3,#4){\includegraphics{#6}} \fi}

\newcommand{\GeV}{\,{\rm GeV}}

\def\vx{\vec{ x}} 
\def\vk{\vec{k}}

\hypersetup{colorlinks, citecolor=bluscuro2, linkcolor=black, urlcolor=bluscuro2}
\definecolor{rossos}{rgb}{0.8,0.2,0.3}
\definecolor{bluscuro1}{rgb}{0.2, 0.2, 0.7}
\definecolor{bluscuro2}{rgb}{0.15, 0.2, 0.9}
\definecolor{verdes}{rgb}{0.1, 0.5, 0.1}

\def\be{\begin{equation}}
\def\ee{\end{equation}}
\def\bea{\begin{eqnarray}}
\def\eea{\end{eqnarray}}

\begin{document}

\color{black}

\begin{flushright} CERN-PH-TH/2012-277\\
SISSA 29/2012/EP\end{flushright}

\vspace{1cm}
\begin{center}
{\Huge\bf\color{black} Non-Gaussianities\\[0.5cm] from the Standard Model Higgs}\\
\bigskip\color{black}\vspace{1cm}{
{\large{Andrea De Simone}$^{a,b,c}$, Hideki Perrier$^{d}$
and {Antonio Riotto}$^{d}$}
\vspace{0.8cm}
} \\[7mm]
{\em $^a$ {CERN, Theory Division, CH-1211 Geneva 23, Switzerland}}\\
{\em $^b$ {SISSA, Via Bonomea 265, I-34136 Trieste, Italy}}\\
{\em $^c$ {INFN, Sezione di Trieste, I-34136 Trieste, Italy}}\\
{\em $^d$ {University of Geneva, Department of Theoretical Physics \\ and Center for Astroparticle Physics (CAP), \\ 24 quai E. Ansermet, CH-1211 Geneva 4, Switzerland}}\\
\end{center}
\bigskip
\centerline{\large\bf Abstract}
\begin{quote}
We have recently proposed that the Standard Model Higgs might be  responsible for generating the cosmological perturbations of the universe by acting as an isocurvature mode during a de Sitter inflationary stage. In this paper we study the level of non-Gaussianity
in the cosmological perturbations which are inevitably generated due to the non-linearities of the Standard Model 
Higgs potential. 
In particular, for the current central value of the top mass, we find that a future detection of non-Gaussianity would exclude 
the detection of tensor modes by the PLANCK satellite.
\end{quote}

\normalsize



\def\thefootnote{\arabic{footnote}}
\setcounter{footnote}{0}
\pagestyle{empty}

\newpage
\pagestyle{plain}
\setcounter{page}{1}

\section{Introduction}
The CERN collaborations  ATLAS \cite{ATLAS} and CMS \cite{CMS} have recently reported the discovery of a boson,  with  mass 
around $(125 -126)$  GeV, whose branching ratios are  fully consistent with   
the Standard Model (SM) Higgs boson. Motivated by this achievement, some of us
 have recently pointed out \cite{us} that the SM Higgs field might be  responsible for generating the cosmological perturbations of the universe by acting as an isocurvature mode during a de Sitter inflationary stage \cite{lrreview}. 
 
 This idea is based 
on the fact that during a period of exponential acceleration with  Hubble rate $H$
all scalar fields
 with a mass smaller than $H$  are inevitably quantum-mechanically excited
with a final superhorizon  flat spectrum. The  comoving curvature perturbation, which provides the initial conditions
for the Cosmic Microwave Background (CMB) anisotropies and for the Large Scale Structure (LSS) of the universe, may be therefore
 generated on superhorizon scales when the isocurvature perturbation, which is  associated to the fluctuations of these light scalar fields,  is converted
into curvature perturbation after (or at the end) of inflation \cite{curvaton1,LW,curvaton3,rate,end,during}. 
In the  inhomogeneous decay rate scenario \cite{rate}  the field responsible for  inflation, the inflaton,  decays  perturbatively with a decay rate 
$\Gamma$. The hot plasma generated by the inflaton decay relics thermalizes with a reheating temperature $T_r\sim (M_{\rm Pl} \Gamma)^{1/2}$. As a consequence, if $\Gamma$  depends on some light fields having a flat spectrum   during inflation,  the decay rate will be characterized by 
 large-scale spatial fluctuations, thus leading to a temperature anisotropy, $\delta T_r/T_r\sim \delta\Gamma/\Gamma$.

In Ref. \cite{us} we have therefore assumed that there was an inflationary period of accelerated expansion during the primordial
evolutionary stage of the universe and that the isocurvature perturbations are
generated by  the SM Higgs field. 
This  allowed us to play with two independent parameters, the SM Higgs mass $m_h$ and the Hubble rate $H$. 
Our findings show that, the Higgs can generate the cosmological perturbation in a wide range of values of the 
Hubble rate during inflation; depending on the values of the SM parameters, e.g. the top-quark mass, and for
the Higgs masses indicated by the recent experimental discovery, we found $H=(10^{10}- 10^{14})$ GeV. 
 On the other side, if the forthcoming data from the Planck satellite will provide 
hints of a $B$-mode in the CMB polarization originated from tensor modes, this can be recast into  a well-defined range of the Higgs mass
\be
(m_h)^{\textrm{B-mode}}\simeq 128.0 \GeV+
1.3\left(\frac{m_t-173.1 \GeV}{0.7 \GeV}\right)  \GeV + 0.9 \left(\frac{H}{10^{15} \GeV}\right) \GeV,
\label{mhfit}
\ee
establishing  a very interesting correlation between collider and cosmological measurements. 

In this paper we further study the idea that the cosmological perturbations may be ascribed to the SM Higgs and focus our attention on another interesting observable characterizing the perturbations, the so-called non-Gaussianity (NG) \cite{reviewNG}.
Primordial NG in  the cosmological perturbations is  a key target for current and future   observational probes both in  the CMB anisotropies and in LSS. The reason is simply stated:  detecting some level of NG  in  the primordial fluctuations as well as its shape allows to  discriminate among different scenarios for the generation of the 
primordial perturbations~\cite{reviewNG}. This is because the different mechanisms giving rise to the 
inflationary perturbations are correlated with  specific  
  shapes.  Models 
such as the one we investigate in this paper, where the curvature perturbation is sourced by some isocurvature perturbation, 
develop  the non-linearities when the perturbation is already on super-Hubble  scales. The  resulting  NG   is a local function of the Gaussian part and therefore is dubbed local. As a consequence, the three-point function in Fourier space gets its major contribution from the ÒsqueezedÓ configuration where one momentum is negligible compared to the others ($k_1 \ll k_2 \sim k_3$). The squeezed limit of NG  leads to strong effects  on the clustering of dark matter halos as the halo bias becomes   strongly scale-dependent  \cite{dalal}. 
 A detection  of primordial NG in the squeezed shape will allow us to    rule out all standard single-field models of 
inflation since they all predict very tiny deviations from Gaussianity \cite{noi,Maldacena}.

In this paper we will answer a very simple question: what is the predicted level of NG in the cosmological perturbations if 
the latter are generated by the SM Higgs? Since we do not wish to restrict ourselves to any specific  mechanism to convert the 
isocurvature perturbation of the SM Higgs into curvature perturbation, we will compute the unavoidable NG generated by the non-linearities
of the SM Higgs quartic potential. This choice is justified  by the fact that   the amount of NG depending on the specific mechanism
of conversion from isocurvature to curvature perturbation may be negligible. We will justify this statement with  a couple of examples.

The paper is organized as follows. In section \ref{sec:perturbation} we give a short summary of primordial NG and how to calculate it. In section \ref{sec:higgsNG} we turn our attention to the NG generated by the non-linear interactions of the SM Higgs.
In section \ref{results} we present our results and draw our conclusions.

\section{Some issues about Non-Gaussianities}
\label{sec:perturbation}
In this paper we will adopt  the $\delta N$ formalism \cite{deltaN}    to study the curvature perturbation. This formalism makes use of the fact that the local expansion  in separate regions  on sufficiently large scales is identified with  the expansion of the unperturbed Friedmann. In general, the curvature perturbation is given by 
\be
\zeta(t_{\rm f},\vec{x})=\delta N+\frac{1}{3}
\int_{\rho(t_*)}^{\rho(t_{\rm f},\vec{x})}\,
\frac{{\rm d}\rho}{\rho+P}
\label{deltaNNNN}
\ee
where $\delta N$ must be interpreted as the  amount of expansion along the worldline of a comoving observer from a spatially
flat  slice at time $t_*$ to a generic slice at time $t_{\rm f}$, and $\rho$ and $P$ are the local energy and pressure densities, respectively. If the  time $t_{\rm f}$ identifies a hypersurface of uniform energy density,  the curvature perturbation at some time $t_{\rm f}$ can be expressed as  a function  of  the values of the relevant light scalar
fields (the inflaton field, the SM Higgs, etc.) $\sigma^I(t_{*},\vec{x})$ at some time $t_{*}$
\be
\zeta(t_{\rm f},\vec{x})=N_I\sigma^I+\frac{1}{2}N_{IJ}\sigma^I\sigma^J+\cdots \,\,\,\, \,\,\,\, \,\,\,\,\,\,\,\, \,\,\,\, \,\,\,\,\,\,\,\, (I=1,\cdots, M),
\label{deltan}
\ee
where $N_I$ and $N_{IJ}$ are the first and second derivative, respectively, of the number of e-folds
with respect to the 
field $\sigma^I$. The corresponding two-point  
correlator of the comoving curvature perturbation is 
\begin{eqnarray}
P_\zeta(k_1)&=&N_IN_JP^{IJ}_{\vec{k}_{1}}, \nonumber\\
\langle\sigma_{\vec{k}_{1}}^I\sigma^J_{\vec{k}_{2}}\rangle&=&(2\pi)^3\delta({\vec{k}_{1}}+{\vec{k}_{2}})P^{IJ}_{\vec{k}_{1}}=
(2\pi)^3\delta({\vec{k}_{1}}+{\vec{k}_{2}})\delta^{IJ}\left(\frac{H}{2\pi}\right)^2
\,.
\end{eqnarray}
Notice that, because of the spatial conformal symmetry  of  the de Sitter geometry, the two-point 
correlators are diagonal in field space \cite{kr}. As we are interested in perturbation generated by the SM Higgs, we will take  $M=2$:  one field is  inflaton $\phi$ and the other the  SM Higgs, $N_\phi\ll N_h$, that is the perturbations are dominated by the SM Higgs.

The three- and four-point correlators of the comoving curvature perturbation, called also  the  bispectrum and trispectrum respectively,  can be also obtained from the expansion (\ref{deltan})
\begin{eqnarray}
\langle \zeta_{\vk_1}\zeta_{\vk_2}
\zeta_{\vk_3}\rangle&=&B^{\rm n-un}_\zeta(\vk_1,\vk_2,\vk_3)+B^{\rm un}_\zeta(\vk_1,\vk_2,\vk_3),\nonumber\\
B^{\rm n-un}_\zeta(\vk_1,\vk_2,\vk_3)&=&N_I N_J N_K B^{IJK}_{\vec{k}_1\vec{k}_2\vec{k}_3},\nonumber\\
B^{\rm un}_\zeta(\vk_1,\vk_2,\vk_3)&=& N_I N_{JK}N_{L}\left(P^{IK}_{\vec{k}_1}P^{JL}_{\vec{k}_2}+2\,\,{\rm permutations} \label{zeta3}
\right)
\end{eqnarray}
and 
\begin{eqnarray}
\langle \zeta_{\vk_1}\zeta_{\vk_2}
\zeta_{\vk_3}\zeta_{\vk_4}\rangle&=&T^{\rm n-un}_\zeta(\vk_1,\vk_2,\vk_3,\vk_4)+T^{\rm un}_\zeta(\vk_1,\vk_2,\vk_3,\vk_4),\nonumber\\
T^{\rm n-un}_\zeta(\vk_1,\vk_2,\vk_3,\vk_4)&=&N_I N_J N_K N_L T^{IJKL}_{\vec{k}_1\vec{k}_2\vec{k}_3\vec{k}_4}
+N_{IJ} N_{K}N_{L}N_M\left(P^{IK}_{\vec{k}_1}B^{JLM}_{\vec{k}_{12}\vec{k}_3\vec{k}_4}+11\,\,{\rm permutations}
\right),\nonumber\\
T^{\rm un}_\zeta(\vk_1,\vk_2,\vk_3,\vk_4)&=&N_{IJ} N_{KL}N_{M}N_N\left(P^{IL}_{\vec{k}_{12}}P^{JM}_{\vec{k}_{1}}P^{KN}_{\vec{k}_{3}}
+11\,\,{\rm permutations}
\right)
\nonumber\\
&&+N_{IJK} N_{L}N_{M}N_N\left(P^{IL}_{\vec{k}_{1}}P^{JM}_{\vec{k}_{2}}P^{KN}_{\vec{k}_{3}}
+3\,\,{\rm permutations}
\right),
 \label{zeta4}
\end{eqnarray}
where  $\vec{k}_{ij}=(\vec{k}_{i}+\vec{k}_{j})$. To take into account the fact that the fields can be NG at Hubble crossing, we have defined their three- and four-point correlators 
\begin{eqnarray}
\langle\sigma_{\vec{k}_{1}}^I\sigma^J_{\vec{k}_{2}}\sigma^K_{\vec{k}_{3}}\rangle&=&(2\pi)^3\delta({\vec{k}_{1}}+{\vec{k}_{2}}+{\vec{k}_{3}})B^{IJK}_{\vec{k}_{1}\vec{k}_{2}\vec{k}_{3}},\nonumber\\
\langle\sigma_{\vec{k}_{1}}^I\sigma^J_{\vec{k}_{2}}\sigma^J_{\vec{k}_{3}}\sigma^L_{\vec{k}_{4}}\rangle&=&(2\pi)^3\delta({\vec{k}_{1}}+{\vec{k}_{2}}+{\vec{k}_{3}}+{\vec{k}_{4}})T^{IJKL}_{\vec{k}_{1}\vec{k}_{2}\vec{k}_{3}\vec{k}_{4}}.\label{kro} 
\end{eqnarray}
 We see in full generality that the pieces contributing to the three- and the four-point correlators of $\zeta$  belong to two different groups: the first is proportional to the connected  correlators of the $\sigma^I$ fields; it originates whenever the   fields $\sigma^I$ are intrinsically NG. This is certainly true for the SM Higgs fluctuations as the SM potential is not quadratic. 
We have decided to call it  the non-universal contribution following Ref. \cite{kr2}.  The second universal group originates from  the modes of 
the fluctuations when they are super-Hubble, even if the $\sigma^I$ fields are gaussian. 

In the presence of  NG, one can define the following  non-linear parameters\footnote{The prime denotes correlators without the $(2\pi)^3\delta^{(3)}(\sum_i\vk_i)$ factors. } 
\begin{eqnarray}
f_{\rm NL}&=&\frac{5}{12}\frac{\langle \zeta_{\vk_1}\zeta_{\vk_2}
\zeta_{\vk_3}\rangle^\prime}{P^\zeta_{\vk_1}P^\zeta_{\vk_2}},
\qquad\qquad\qquad\qquad\;\;({\rm squeezed}:\,\,k_1\ll k_2\sim k_3) \label{fnlsqueezed}\\
\tau_{\rm NL}&=&\frac{1}{4}\frac{\langle \zeta_{\vk_1}\zeta_{\vk_2}\zeta_{\vk_3}\zeta_{\vk_4}\rangle^\prime}{P^\zeta_{\vk_1}
P^\zeta_{\vk_3}P^\zeta_{\vk_{12}}},
\qquad\qquad\qquad\qquad({\rm collapsed}:\,\,\vk_{12}\simeq  0), \label{tf2}\\
2\tau_{\rm NL}+\frac{54}{25}g_{\rm NL}&=&\frac{\langle \zeta_{\vk_1}\zeta_{\vk_2}\zeta_{\vk_3}\zeta_{\vk_4}\rangle^\prime}{P^\zeta_{\vk_4}
\left(P^\zeta_{\vk_1}P^\zeta_{\vk_{2}}+2\,\,{\rm permutations }\right)},
\quad({\rm squeezed}:\,\,k_4\ll k_1,k_2,k_3). \label{tf3}
\end{eqnarray}
These parameters get both non-universal and universal contributions. For instance,
\begin{eqnarray}
f_{\rm NL}&=&f^{\rm non-un}_{\rm NL}+f^{\rm un}_{\rm NL},\nonumber\\
f^{\rm non-un}_{\rm NL}&=&\frac{5}{12}\frac{B^{\rm n-un}_\zeta(\vk_1,\vk_2,\vk_3)}{(N_I N^I)^4},\nonumber\\
f^{\rm un}_{\rm NL}&=&\frac{5}{6}\frac{N^IN_{IJ}N^J}{(N_IN^I)^2}.
\end{eqnarray}
In this paper we will focus our attention to the contributions to the NG coming from the non-universal pieces: we can compute them exactly
once the SM Higgs potential is known. The universal pieces are model-dependent, {\it i.e.} they depend on the specific mechanism by which the isocurvature perturbation is converted into the curvature perturbation. We will assume that their contribution to the total NG is subdominant, and motivate this assumption with  a couple of examples in the following two sub-sections. 
Of course, in a given well-defined model one should explicitly check  the validity of this hypothesis.

\subsection{Non-Gaussianities in the modulated decay scenario}
As we mentioned in the Introduction, a specific example where the primordial density perturbations may be produced just
after the end of inflation is the modulated decay scenario when the decay rate of the inflaton is a function of the SM Higgs field \cite{rate}, that is  $\Gamma=\Gamma(h)$. Therefore, from now one we restrict ourselves to the case in which there is only one relevant fluctuating field, the SM Higgs. 
If we approximate the inflaton reheating by a sudden decay, we may find an analytic estimate of the density perturbation.
In the case of modulated reheating, the decay occurs on a spatial hypersurface with
variable local decay rate and hence local Hubble rate $H=\Gamma(h)$. Before the 
inflaton decay, the oscillating inflaton field has a pressureless equation of state and there is no density
perturbation. The perturbed expansion reads
\be
\delta N_{\rm d}=-\frac{1}{3}\ln\left(\frac{\rho_\phi}{\overline{\rho}_\phi}\right).
\ee
As the universe is made of radiation after the inflation decay, we can write
\be
\zeta=\delta N_{\rm d}+\frac{1}{4}\ln\left(\frac{\rho_\phi}{\overline{\rho}_\phi}\right).
\ee
Eliminating $\delta N_{\rm d}$ and using the local Friedmann equation $\rho\sim H^2$, to determine the local density in terms of the local decay rate $\Gamma=\Gamma(h)$, we have at the linear order
\be
\label{rate}
\zeta=-\frac{1}{6}\,\frac{\delta\Gamma}{\Gamma}=
-\frac{1}{6}\,\frac{{\rm d\ln }\Gamma}{{\rm d}\ln \overline{h}}\,\frac{\delta h}{\overline h}= -\beta_h\frac{\delta h}{\overline{h}}\,,
\ee
where $\bar h$ is the vacuum expectation value of the Higgs field during inflation.
The corresponding power spectrum of the curvature perturbation is  given by
\be
\label{aaaa}
P_{\zeta}= \beta_h^2\left(\frac{H}{2\pi\overline{h}}\right)^2.
\ee
By going to higher orders in the fluctuation $h$, we get
(primes indicate differentiation with respect to the
Higgs field)
\begin{eqnarray}
f^{\rm un}_{\rm NL}&=&5\left(1-\frac{\Gamma''\Gamma}{\Gamma'^{2}}\right),\nonumber\\
g^{\rm un}_{\rm NL}&=&\frac{25}{54}\frac{N'''}{N'^{3}}=
\frac{50}{3}\left(2-3\frac{\Gamma''\Gamma}{\Gamma'^{2}}+\frac{\Gamma^2\Gamma'''}{\Gamma'^{3}}\right).
\end{eqnarray}
Now, suppose that the function $\Gamma(h)$ gets its dependence on the SM Higgs from some Yukawa-type interaction with Yukawa coupling $Y=Y_0(1+h/2M)$, with $M$ some high mass scale.
If so, $\Gamma(h)\simeq \Gamma_0(1+h/2M)^2$ and we obtain $f^{\rm un}_{\rm NL}= 5/2$ and 
$g^{\rm un}_{\rm NL}= 25/3$. Both these non-linear parameters  lead to 
a level of NG which is difficult to test observationally and therefore subdominant.
For further discussion about this particular mechanism, see also Ref.~\cite{choi}.

\subsection{Non-Gaussianities generated at the end of inflation}

Another possibility is that that the dominant component of
the curvature perturbation is generated at the transition between inflation and the post-inflationary phase \cite{end}.  
Let us suppose that slow-roll inflation suddenly
gives way to radiation domination through a waterfall transition in hybrid inflation and that the value of the inflaton $\phi_{\rm e}$ at which
this happens depends on the SM Higgs field, $\phi_{\rm e}=\phi_{\rm e}(h)$. If we assume again that the contribution to the
curvature perturbation from the inflaton field is subdominant, we get
\be
\zeta= N'_{\rm e}\delta \phi_{\rm e}=N'_{\rm e}\frac{{\rm d}\phi_{\rm e}}{{\rm d}\ln \overline{h}}\frac{\delta h}{\overline{h}}=
\frac{{\rm d}N_{\rm e}}{{\rm d}\ln \overline{h}}\frac{\delta h}{\overline{h}}.
\ee
For concreteness, let us assume  that inflation ends on those Hubble
regions where $\lambda_\phi \phi^2+ g^2h^2=$ constant. In this case,
the corresponding power spectrum of linear perturbations is given by
\be
\label{ay}
P_{\zeta_{\rm g}}=
\frac{\overline{h}^2}{2M_{\rm Pl}^2\epsilon_{\rm e}}\left(\frac{g^2 \overline{h}}{\lambda_\phi \phi_{\rm e}}\right)^2\left(\frac{H}{2\pi\overline{h}}\right)^2,
\ee
where $\epsilon_{\rm e}=-\dot H/H^2$ at the time of the waterfall transition. The universal contribution to the three-point correlator is given by
(primes again indicate differentiation with respect to the SM Higgs)
\begin{eqnarray}
\frac{6}{5}f^{\rm un}_{\rm NL}&=&\frac{N'_{\rm e}\phi''_{\rm e}+N''_{\rm e}\phi^{'2}_{\rm e}}{(N'_{\rm e}\phi'_{\rm e})^2}\simeq
\frac{\phi''_{\rm e}}{(N'_{\rm e}\phi^{'2}_{\rm e})}+\eta_{\rm e}-2\epsilon_{\rm e}\nonumber\\
&\simeq&-\sqrt{2\epsilon_{\rm e}}\frac{M_{\rm Pl}}{\overline{h}}\frac{\lambda_\phi \phi_{\rm e}}{g^2 \overline{h}}\left(1+\frac{g^4 \overline{h}^2}{\lambda^2_\phi \phi^{2}_{\rm e}}\right),
\end{eqnarray}
where we have introduced the slow-roll parameter $\eta_{\rm e}$ through the relation $N''_{\rm e}=N^{'2}_{\rm e}(\eta_{\rm e}-2\epsilon_{\rm e})$. If the power spectrum induced by the SM Higgs field dominates over the one generated by the inflaton field $P_{\zeta_\phi}=(1/2\epsilon_k M_{\rm Pl}^2)(H/2\pi)^2$, that is 
\be
(g^2\overline{h}/\lambda_\phi\phi_{\rm e})^2\gg (\epsilon_{\rm e}/\epsilon_k)\gg 1\,,
\label{condizione}
\ee
where $\epsilon_k$ is the slow-roll parameter when the wavenumber $k$ is exiting the Hubble radius,  
then
\be
\frac{6}{5}f^{\rm un}_{\rm NL}\simeq -\sqrt{2\epsilon_{\rm e}}\frac{g^2 M_{\rm Pl}}{\lambda_\phi \phi_{\rm e}}\,.
\ee
This value of NG can be small as long as 
 $\sqrt{2\epsilon_{\rm e}} M_{\rm Pl}\lesssim\overline{h}$ and $\epsilon_{\rm e}\ll 1$,
once Eq.~(\ref{condizione}) is taken into account. 
Having shown that the universal contribution to NG may be small, we now focus our attention on the non-universal contribution, which we can compute exactly once the SM Higgs potential is known.

\section{The SM Higgs contribution to the non-Gaussianity}
\label{sec:higgsNG}

Let us now consider the SM Higgs field   $h(\vx,\tau)$ with potential
$V(h)=\lambda(h)h^4/4$.
If the SM Higgs field has a non vanishing vacuum expectation value $\overline{h}$, we know that
the $n$-point correlator is given by 
\be
\langle h_{\vk_1}(\tau)h _{\vk_2}(\tau)\cdots h_{\vk_n}(\tau)\rangle
=-i\Big< 0\left|\int_{-\infty}^\tau\,{\rm d}\tau'\,\left[h_{\vk_1}(\tau) h_{\vk_2}(\tau)\cdots h_{\vk_n}(\tau),V(h(\tau'))
\right] \right|0\Big>.
\ee
Using the mode functions for a massless field in de Sitter
\be
h_{\vk}(\tau)=\frac{H}{\sqrt{2k^3}}\left(1-ik\tau\right)e^{ik\tau},
\ee
one obtains \cite{zaldarriaga}
\be
\langle h_{\vk_1} h_{\vk_2}\cdots h_{\vk_n}\rangle'=V^{(n)}H^{2n-4} \frac{
(k^{(n)}_t)^3}{\prod_{i=1}^n 2 k_i^3}\, I_n(k_1,k_2,\cdots,k_n),  \label{n-pt}
\ee
where 
$k^{(n)}_t=k_1+k_2+\cdots k_n$, $V^{(n)}={\rm d}^n  V(\overline{h})/{\rm d}\overline{h}^n$ and 
\be
I_n(k_1,k_2,\cdots,k_n)=2\int_{-\infty}^{\tau_{\rm end}} \frac{{\rm d}\tau}{k_t^3\tau^4}{\rm{Re}}\Big{[}-i(1-ik_1 \tau)\cdots(1-ik_n\tau)e^{ik_t^{(n)}\tau}\Big{]}.
\ee
In particular, we have
\begin{eqnarray}
I_3(k_1,k_2,k_3)&=&\frac{8}{9}-\frac{\sum_{i<j} 2 k_i k_j}{\left(k^{(n)}_t\right)^2}-\frac{1}{3}\left(\gamma_{\rm E}+N_{k_t}\right)\frac{\sum_{i} 2 k^3_i}{\left(k^{(n)}_t\right)^3},\,\nonumber\\
I_4(k_1,k_2,k_3,k_4)&=&\frac{8}{9}-\frac{\sum_{i<j} 2 k_i k_j}{\left(k^{(n)}_t\right)^2}+2\frac{\Pi_{i}  k_i }{k_t^4}
-\frac{1}{3}\left(\gamma_{\rm E}+N_{k_t}\right)\frac{\sum_{i} 2 k^3_i}{\left(k^{(n)}_t\right)^3}.
\end{eqnarray}
Here  $\gamma_{\rm E}$ denotes the Euler gamma, while $N_{k_t}$ is the number of e-folds from the time 
the mode $k_t$ crosses the Hubble radius to the time $\tau_{\rm end}$ when inflation ends.

How large can the non-universal contributions be? Let us consider the three-point correlator in the squeezed limit
\be
\langle h_{\vk_1} h_{\vk_2} h_{\vk_3}\rangle'\simeq-\frac{2}{3} \cdot\frac{6\lambda\overline{h}}{H^{2}} \, N_{k_t}\, P_{\vk_1}P_{\vk_2},
\qquad\qquad\quad(k_1\ll k_2\sim k_3).
\ee  
The corresponding contribution to the three-point correlator of the comoving curvature perturbation is 

\be
\langle\zeta_{\vk_1}\zeta_{\vk_2} \zeta_{\vk_3}\rangle'\simeq-4 
\frac{\lambda\overline{h}}{H^{2}}  \, \frac{N_{k_t}}{N^{\prime }}\, P^\zeta_{\vk_1}P^\zeta_{\vk_2},
\qquad\qquad\quad(k_1\ll k_2\sim k_3)
\ee  
leading to a non-universal contribution to the non-linear parameter $f_{\rm NL}$ given by
(see Eq.~(\ref{fnlsqueezed}))
\begin{eqnarray}
f_{\rm NL}^{\rm n-un}\simeq -\frac{5}{3} \frac{\lambda\overline{h}}{H^{2}} \frac{N_{k_t}}{N^{\prime}}&=&-\frac{5 }{6\pi} \frac{\lambda\overline{h}}{H} \frac{N_{k_t}}{{\cal P}_\zeta^{1/2}}
\simeq 
-5.0 \left(\frac{\lambda}{10^{-3}}\right)^{1/2}\left(\frac{m_h(\overline{h})/H}{10^{-3}}\right) \left(\frac{N_{k_t}}{50}\right)\,,
\label{fNLresult}
\end{eqnarray}
where we have defined the quantity ${\cal P}_\zeta\equiv (k^3/2 \pi^2)P^\zeta_{\vk}=N^{\prime 2}(H/2\pi)^2\simeq 2.3\times  
10^{-9}$ and the (field-dependent) Higgs mass $m^2_h(\overline{h})=3\lambda\overline{h}^2$. 
Analogously we find
the  four-point correlator to be
\begin{eqnarray}
\langle h_{\vk_1}h_{\vk_2}h_{\vk_3} h_{\vk_4}\rangle'&=&6\lambda H^4\, \frac{(k^{(4)}_t)^3}{16\,k_1^3\,k_2^3\,k_3^3\,k_4^3}I_4(k_1,k_2,k_3,k_4), 
\end{eqnarray}
which, in the squeezed limit $k_4\ll k_1,k_2,k_3$,    reduces to 
\begin{eqnarray}
\langle h_{\vk_1}h_{\vk_2}h_{\vk_3} h_{\vk_4}\rangle'&=&6\lambda H^4\, \frac{(k^{(3)}_t)^3}{16\,k_1^3\,k_2^3\,k_3^3\,k_4^3}I_3(k_1,k_2,k_3)
=\frac{1}{{\overline{h}}}\, \frac{H^2}{2\,k_4^3}\langle h_{\vk_1}h_{\vk_2}h_{\vk_3}\rangle'\nonumber\\
&=& \frac{P_{\vk_4}}{\overline{h}}\langle h_{\vk_1} h_{\vk_2} h_{\vk_3}\rangle'.
\end{eqnarray}
This corresponds to
\begin{eqnarray}
g_{\rm NL}^{\rm n-un}\simeq -\frac{25}{54} \frac{2\lambda}{H^{2}} \frac{N_{k_t}}{N^{\prime 2 }}=
-\frac{25}{27(2\pi)^2} \frac{\lambda N_{k_t}}{{\cal P}_\zeta}\simeq -5.1\times 10^5\left(\frac{\lambda}{10^{-3}}\right) \left(\frac{N_{k_t}}{50}\right)\,.
\label{gNLresult}
\end{eqnarray}
Notice that both $f_{\rm NL}$ and $g_{\rm NL}$ are predicted to be negative.

\section{Results and conclusions}
\label{results}

\begin{figure}[t]
\centering
\includegraphics[scale=0.8]{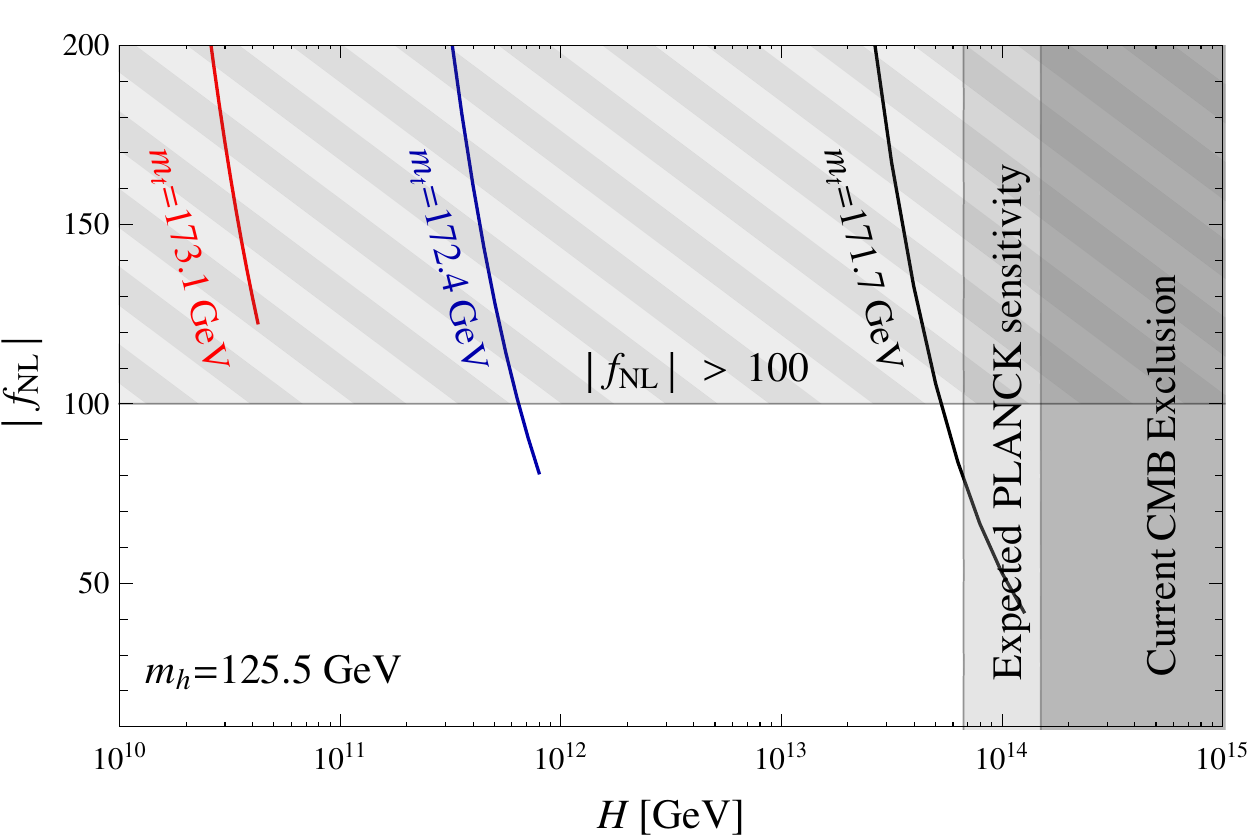}
\caption{{\small\textit{ 
The absolute value of $f_{\rm NL}$ as a function of $H$, for $m_h=125.5 \GeV$ and
three values of the top mass: $m_t=171.7, 172.4, 173.1 \GeV$.
The curves correspond to values of $H$ below the instabiity scale, and they end where 
$H=\Lambda_{\rm inst}$.
The limit $|f_{\rm NL}|<10^2$ has also been displayed as horizontal band.
}}}
\label{fig:fNLvsH}
\end{figure}

To find the amount of NG generated by the SM Higgs we need to compute its potential
and the solution to the field equation of motion in the expanding background.
As in Ref.~\cite{us}, 
 we adopt 2-loop renormalization group (RG) equations for 
gauge, Higgs-quartic and top-yukawa couplings;
 the pole-mass matching scheme for the Higgs and top masses is taken from 
 the Appendix of Ref.~\cite{hi3}.
We considered $m_t=173.1\pm 0.7$ GeV  for the top mass \cite{hi5}, 
and $\alpha_s(M_Z)=0.1184\pm 0.0007$ 
 for the QCD gauge coupling \cite{alphas}.
For simplicity,  we have only considered the central value of the QCD coupling $\alpha_s$;
in fact, the size of the effect of the $1\sigma$-variation of $\alpha_s$ 
turns out to be comparable with the higher-order effects that we are neglecting (e.g.~three-loops effects).
The RG-improved effective potential $V(h)$ of the SM Higgs \cite{2loop} is then obtained by solving
numerically the RG equations, as a function of input parameters, such as the top and Higgs masses.

The dynamics of the SM Higgs field during inflation is found, 
after setting the Hubble rate during inflation and the initial value of the  field, by solving the equation
\be
\ddot h+3 H \dot h+ V'(h)=0\,,
\label{eq:fieldeq}
\ee
where the dot refers to derivative with respect to $t=H^{-1}\ln a$ (conventionally, the scale factor is set to 1 
at the initial field value). 
The dynamics has to satisfy the following requirements: 1)  
$V(h)\ll H^2 M_{\rm Pl}^2$, which means that 
the contribution of the Higgs field  to the energy density during inflation is negligible;
2) $|{\rm d}^2 V(h)/{\rm d} h^2|\ll H^2$, which means that 
the SM Higgs field is light enough during inflation.
The latter condition must hold for enough e-folds to ensure sufficient homogeneity and isotropy 
of the present observable universe (we consider 60 e-folds as a fiducial number).
An automatic consequence of the condition 2 is that 
 the special index of the perturbations is sufficiently close to 1.

\begin{figure}[t]
\centering
\includegraphics[scale=0.8]{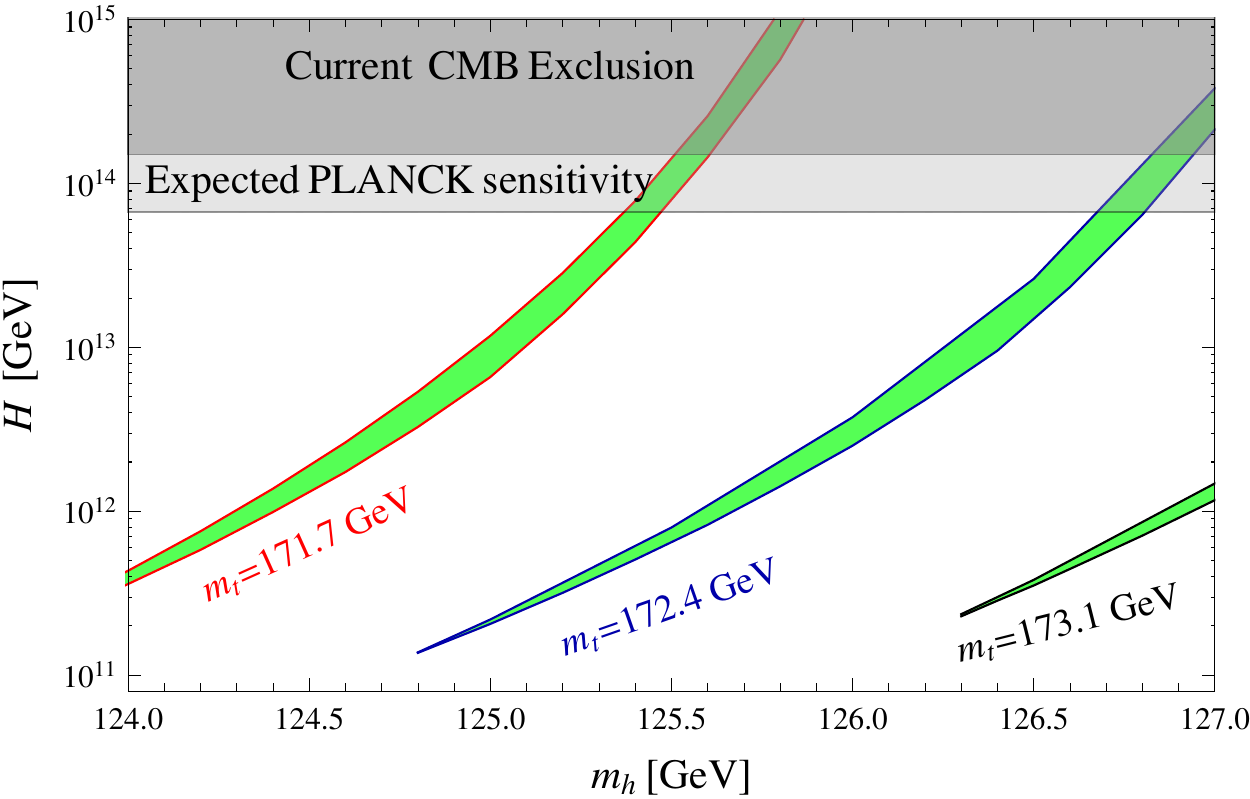}
\caption{{\small \textit{The green bands indicate the allowed values of $m_h, H$, where $H<\Lambda_{\rm inst}$ and 
$|f_{\rm NL}|<100$,
for different top-quark masses (the central value, $1\sigma$ and $2\sigma$ below). }}
}
\label{fig1}
\end{figure}

\begin{figure}[t]
\centering
\includegraphics[scale=0.8]{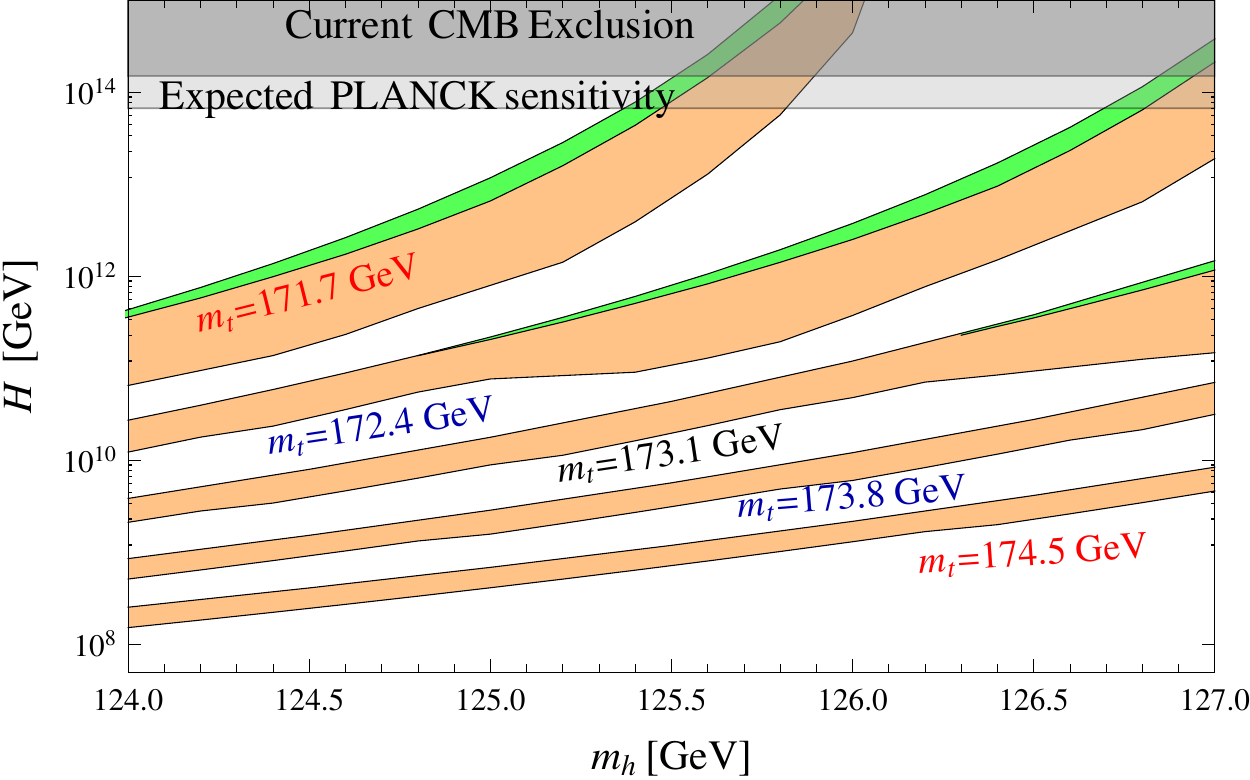}
\caption{{\small\textit{ 
The orange bands indicate the allowed values of $m_h, H$, where $H<\Lambda_{\rm inst}$ and 
$|g_{\rm NL}|<10^6$,
for different top-quark masses (the central value, $\pm1\sigma$ and $\pm2\sigma$).
The green bands correspond to the regions allowed by the $f_{\rm NL}$ constraint, as in Fig.~\ref{fig1}.}}
}
\label{fig2}
\end{figure}

Next, we  scan over
the Higgs and top masses and the Hubble rate 
 $H$, looking for those values for which the cosmological perturbations have the correct value,  $P_\zeta^{1/2}=4.8\times 10^{-5}$, 
and computing the amount of  NG generated according to Eqs.~(\ref{fNLresult}) and (\ref{gNLresult}).
Of course, we have to impose that the NG are sufficiently small to meet the current bounds, $|f_{\rm NL}|\lesssim 10^2$ \cite{kom} and $|g_{\rm NL}|\lesssim
 10^6$ \cite{gnl}. 
 
 The instability scale
$\Lambda_{\rm inst}$ of the SM Higgs potential at which the Higgs quartic coupling runs negative sets
an upper limit on the values of $H$.
A lower limit on $H$ is set by the NG constraints.
The way to keep $f_{\rm NL}$ below the bound is by having  $\lambda(\bar h)$ small and this is achieved by
taking $\bar h$ as close as possible to $\Lambda_{\rm inst}$, where $\lambda=0$. 
The field can roll down starting, at most, from the maximum of the potential, and this is our 
 initial condition for the equation of motion (\ref{eq:fieldeq}).

 Our results are presented in Figs.~\ref{fig:fNLvsH}, \ref{fig1} and \ref{fig2}  where we also show the current exclusion
  limit on $H$ from CMB data which dictates $H\lesssim 1.5 \times 10^{14}$ GeV \cite{wmapping} (darker gray band), 
and the testable region by the PLANCK experiment
$6.7 \times 10^{13} \GeV \lesssim H \lesssim 1.5 \times 10^{14} \GeV$  \cite{plancktensor} (lighter gray band).

In Figure \ref{fig:fNLvsH}, we plot the values of $f_{\rm NL}$, as a function of $H$, for fixed $m_h=125.5 \GeV$ and
different values of the top mass: $m_t=171.7, 172.4, 173.1 \GeV$.
The curves correspond to values of $H$ below the instability scale, and they end where 
$H=\Lambda_{\rm inst}$.
The limit $|f_{\rm NL}|<10^2$ has also been displayed as a horizontal band.
Notice that, for a given $m_h$, a sufficently small amount of NG is produced as long as the top
mass is small enough. This because the instability scale increases for smaller $m_t$, and
this  allows larger values of $H$ to be considered, and thus smaller $f_{\rm NL}$.

The bands in Fig.~\ref{fig1} correspond to the upper and lower limits on $H$ discussed above, 
for different values of the Higgs and  top-quark masses.
In particular, for the central value of $m_t=173.1$ GeV, $H$ is limited  to be in the range $(10^{11} - 10^{12}) \GeV$,
and therefore no tensor modes would be seen.
Notice that the $f_{\rm NL}$ constraint implies a lower bound on $H$ almost  independent on $m_h, m_t$:
 $H\gtrsim 10^{11}$ GeV. This can be recovered by imposing $f_{\rm NL}=-100$ in Eq.~(\ref{fNLresult}).
 In the range of masses we considered we typically have  $\bar h\gtrsim 10^{11}$ GeV and 
 $\lambda \gtrsim 4 \times10^{-4}$. This implies a minimum value of $H\sim 10^{11} \GeV$.

In Fig.~\ref{fig2},  the corresponding limits from $g_{\rm NL}$ are shown (orange bands).
We deduce that whenever $f_{\rm NL}$ is large, $g_{\rm ¥NL}$ is small 
and therefore the bounds on the bispectrum are stronger the the ones coming from the trispectrum.

One can also obtain a fitting formula 
relating the Higgs mass, the top mass and the Hubble rate on the isocurve $f_{\rm NL}=-100$,
\be
(m_{h})^{f_{\rm NL}=-100}\simeq 128.1\GeV +1.3\left(\frac{m_t-173.1 \GeV}{0.7 \GeV}\right)\GeV +1.3\left(\frac{H}{10^{15} \GeV}\right)\GeV .
\label{fit}
\ee
This formula is valid in the ranges 124.0 GeV $\lesssim m_h \lesssim 127.0$ GeV and 
$6.7 \times 10^{13}\GeV \lesssim H \lesssim 1.5 \times 10^{14}$ GeV.
If tensor modes will be detected with a given $H$, Eq.~(\ref{fit}) makes a prediction for the Higgs and top masses
corresponding to a level of NG of $f_{\rm NL}=-100$.

In conclusion, we have studied the level of NG in the cosmological perturbations which are generated by
the non-linearities of the SM Higgs potential under the hypothesis that the SM Higgs is responsible for the generation
of the curvature perturbation (but not for   driving inflation). In particular, we have assumed that the universal contribution
to the NG is negligible, supporting this hypothesis with a couple of examples. In this sense, the NG studied in this paper 
is  unavoidable (barring potential cancellations with the universal contributions).
Under these circumstances we have obtained that:
\begin{itemize}
 \item The NG in the four-point correlator (in the squeezed limit) is always negligible once the bounds on the amplitude of the three-point correlator are accounted for.
 
 \item The current constraints on $f_{\rm NL}$ imply a lower bound on the Hubble rate during inflation, $H\gtrsim 10^{11}$ GeV, in the range of Higgs mass indicated by the LHC experiments.
 
 \item  For the current central value of the top mass ($m_t=173.1 \GeV$) a future detection of NG would exclude the detection of tensor modes through the $B$-mode of the CMB polarization.

\end{itemize}

\section*{Acknowledgments}
H.P. and A.R. are supported by the Swiss National
Science Foundation (SNSF), project ``The non-Gaussian Universe" (project number: 200021140236).

\begin{small}

\end{small}
\end{document}